\documentclass[lettersize,journal]{IEEEtran}
\usepackage{amsmath,amsfonts}
\usepackage{algorithmic}
\usepackage{algorithm}
\usepackage{array}
\usepackage[caption=false,font=normalsize,labelfont=sf,textfont=sf]{subfig}
\usepackage{textcomp}
\usepackage{stfloats}
\usepackage{url}
\usepackage{verbatim}
\usepackage{graphicx}
\usepackage{cite}
\begin{document}

\title{Microwave electrometry with multi-photon coherence in Rydberg atoms}

\author{Zheng Yin, Shengfang Zhao, Yuan He, Zhengmao Jia, Michal Parniak, Xiao Lu and Yandong Peng
\thanks{This work was supported by the Shandong Natural Science Foundation, China ZR2021LLZ006, and the Taishan Scholars Program of Shandong Province, China ts20190936.}
\thanks{Zheng Yin is with the College of Electronic and Information Engineering, and also with the College of Electrical Engineering and Automation, Shandong University of Science and Technology, Qingdao 266590, China.}
\thanks{Shengfang Zhao, Yuan He and Zhengmao Jia are with the College of Electronic and Information Engineering, Shandong University of Science and Technology, Qingdao 266590, China.}
\thanks{Michal Parniak is the Centre for Quantum Optical Technologies, Centre of New Technologies, University of Warsaw, Banacha 2c, 02-097 Warsaw, Poland (e-mail: m.parniak@cent.uw.edu.pl).}
\thanks{Xiao Lu is with the College of Electrical Engineering and Automation, Shandong University of Science and Technology, Qingdao 266590, China.}
\thanks{Yandong Peng is with the College of Electronic and Information Engineering, Shandong University of Science and Technology, Qingdao 266590, China (e-mail: pengyd@sdust.edu.cn).}}

\markboth{IEEE PHOTONICS JOURNAL,~Vol.~xx, No.~xx, xxxx~2023}%
{Shell \MakeLowercase{\textit{et al.}}: A Sample Article Using IEEEtran.cls for IEEE Journals}


\maketitle

\begin{abstract}
A scheme for measurement of microwave (MW) electric field is proposed via multi-photon coherence in Rydberg atoms. It is based on the three-photon electromagnetically induced absorption (TPEIA) spectrum. In this process, the multi-photon produces a narrow absorption peak, which has a larger magnitude than the electromagnetically induced transparency (EIT) peak under the same conditions. The TPEIA peak is sensitive to MW fields, and can be used to measure MW electric field strength. It is interesting to find that the magnitude of TPEIA peaks shows a linear relationship with the MW field strength. The simulation results show that the minimum detectable strength of the MW fields is about 1/10 that based on an common EIT effect, and the probe sensitivity is improved by about 4 times. Furthermore, the MW sensing based on three-photon coherence shows a broad tunability, and the scheme may be useful for designing novel MW sensing devices.
\end{abstract}


\section{Introduction}
\IEEEPARstart{A}{tom}-based metrology has been widely used in many fields, such as atomic clocks, measurement of temperature, frequency, magnetic and electric fields, due to the unique properties of atoms and molecules \cite{ludlow2015optical,kucsko2013nanometre,sun2017measuring,kominis2003subfemtotesla,sedlacek2012microwave}. Recently, Rydberg-atom-based microwave electrometry arises great interest \cite{osterwalder1999using,holloway2014broadband,sedlacek2012microwave}. Rydberg atoms with one or more electrons of large principal quantum numbers are sensitive to electric fields, and can coherently interact with a microwave (MW) electric field. It can significantly increase the accuracy and repeatability of measurement. The main research works are based on electromagnetically induced transparency (EIT) and Aulter-Towns splitting \cite{gallagher2005rydberg,boller1991observation,alzetta1997induced}, where two laser fields drove atoms to their Rydberg states and Rydberg EIT splitting induced by a MW field was used for MW electrometry \cite{sedlacek2012microwave,cheng2017high,simons2016using,fan2015atom,sedlacek2013atom}. New achievements about MW measurement include Rydberg-atom-based superheterodyne receiver \cite{jing2020atomic}, enhanced MW metrology by population repumping \cite{prajapati2021enhancement}, broadband terahertz wave detection \cite{zhou2022theoretical}, arrival angle of microwave signals \cite{tu2017angle,robinson2021determining}, continuous measurement \cite{simons2021continuous} and auxiliary transition \cite{jia2021span}, and radio-frequency phase measurement \cite{simons2019rydberg,lin2022sensitive}, etc.

 It is know that an EIT system has good coherence and its probe spectrum is widely used for MW measurement. Recently, three-photon coherence attracts researchers' interest. For example, observation of three-photon electromagnetically induced absorption (TPEIA) \cite{moon2014three} in atomic systems, constructive interference in the three-photon absorption \cite{mcgloin2001electromagnetically,carr2012three}, demonstration of three-photon coherence condition \cite{lee2015relationship}, and its extention to Rydberg atoms \cite{kwak2016microwave}. Three-photon coherence has important applications, such as overcoming residual Doppler shifts \cite{shaffer2018read},  miniaturizing of MW field sensors and receivers \cite{you2022microwave} and its dependence on the AT splitting with two coupling lasers \cite{bai2022autler}. While, few reaserch involves MW measurement based on three-photon coherence.
 
 In this paper, we propose a scheme to measure MW electric fields based on three-photon coherence in Rydberg atoms. A probe and a control fields counter-propagate through the atomic system \cite{gutekunst2017fiber}. The ${}^{87}{\rm{Rb}}$ atoms are excited from the ground states to the Rydberg states, and the absorption spectrum of three-photon transition shows a single absorption peak around the resonant frequency. Due to three-photon coherence, a strong TPEIA peak appears under the three-photon resonance condition. It is interesting to find that the TPEIA peak changes linearly with the MW electric field strength. This may be used to detect the MW electric field. The simulation results show the sensitivity is enhanced by about 4 times and its minimum detectable strength of the MW electric field can be increased by more than one order of magnitude, compared with a common EIT scheme. Also, the scheme shows a wide tunability. This may help to design novel MW sensing devices.
\begin{figure*}[!t]
\centering
\subfloat[]{\includegraphics[width=4.9cm,height = 4.3cm]{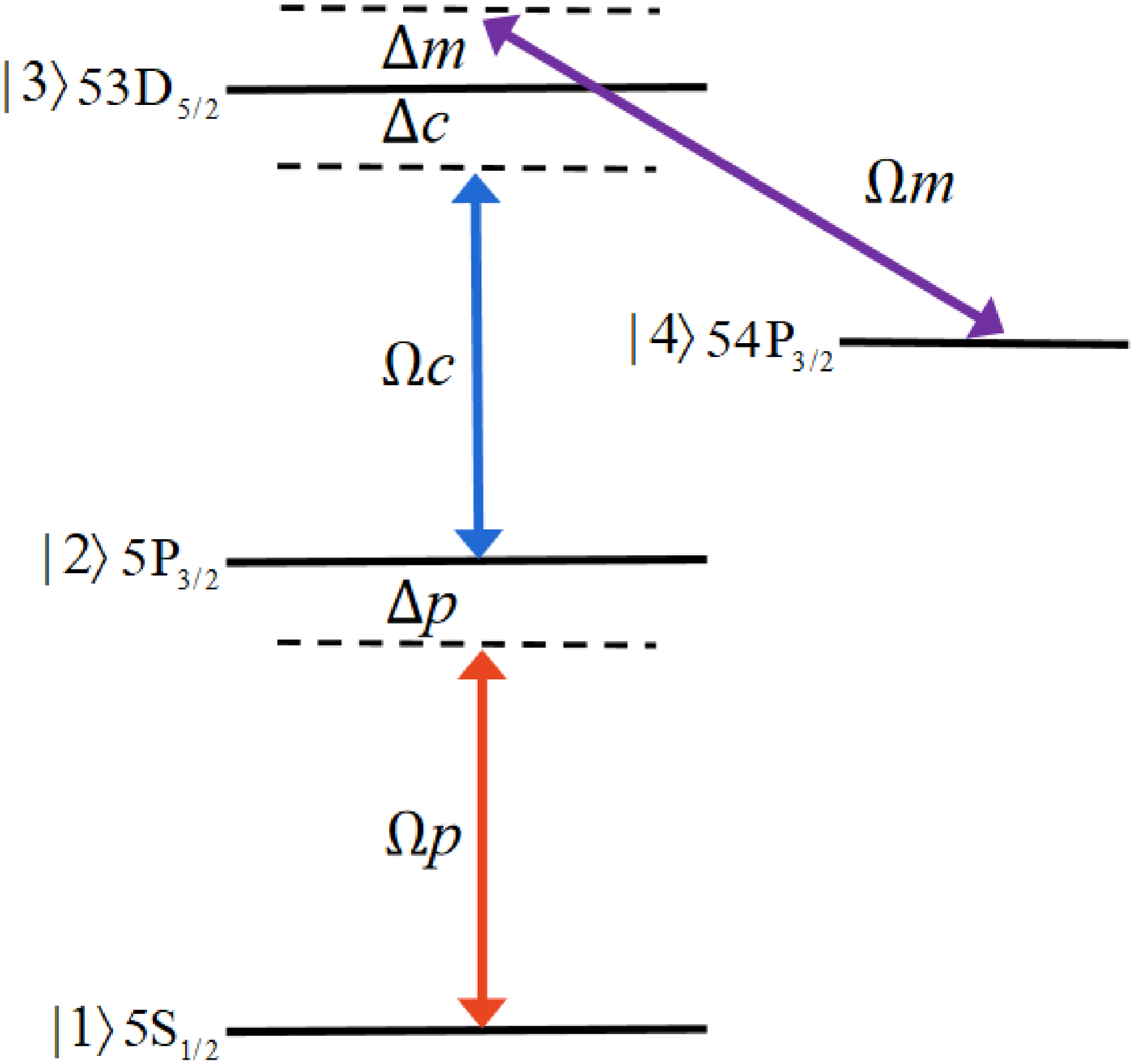}%
\label{Energy level diagramSchematic}}
\hfil
\subfloat[]{\includegraphics[width=6.5cm,height = 4.3cm]{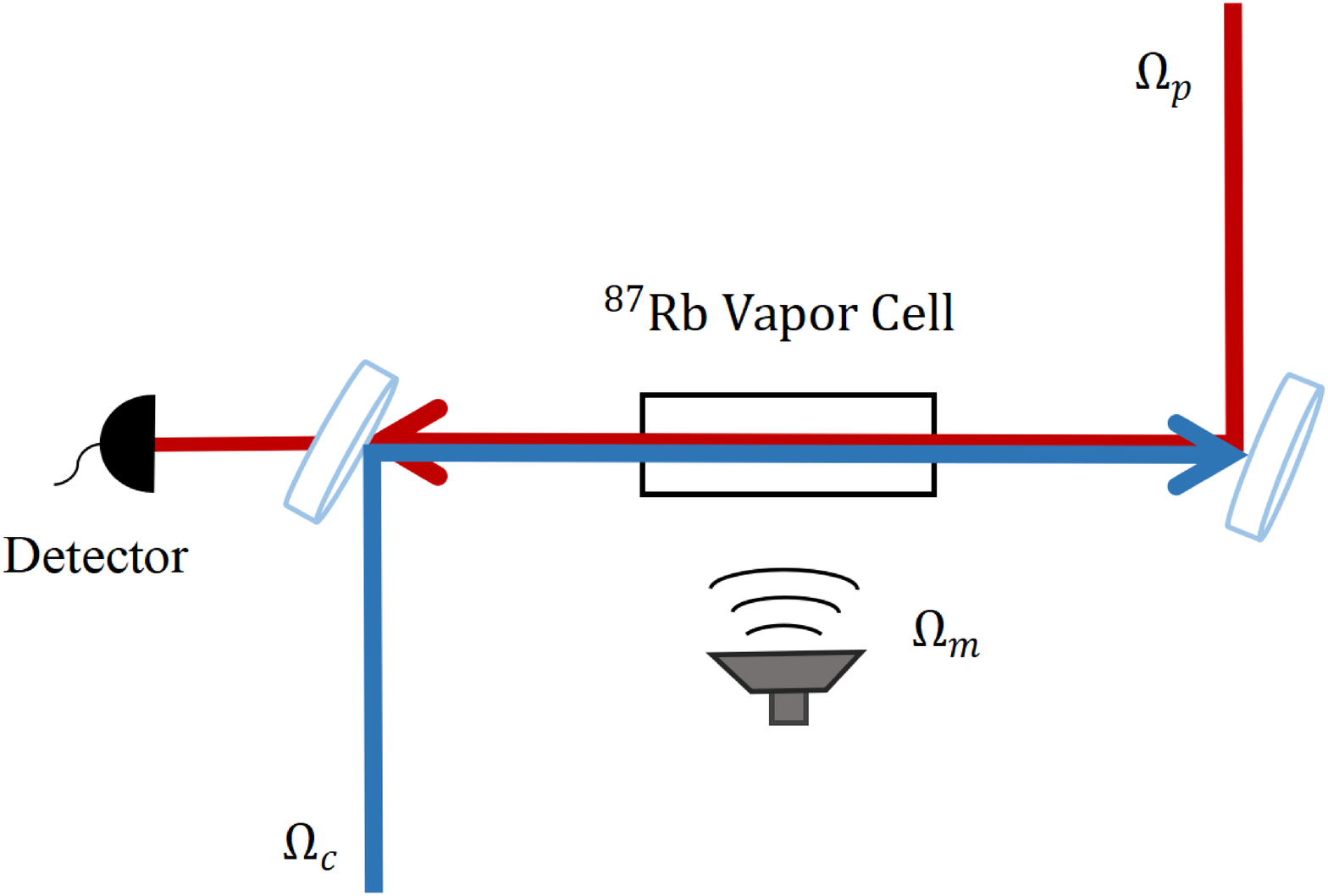}%
\label{Schematic}}
\caption{(a) Four-level Rydberg atom model and (b) schematic diagram of experimental setup including an atoms cell, probe (${\Omega _p}$), coupling (${\Omega _c}$), a microwave signal generator and detector. }
\label{Figure. 1}
\end{figure*}

\section{MODEL AND BASIC EQUATION}
Fig. 1(a) shows a four-level ladder-type atomic system. The relevant atomic energy levels of ${}^{87}{\rm{Rb}}$ are $5{S_{1/2}}$ ($|1\rangle $) , $5{P_{3/2}}$ ($|2\rangle $), $53{D_{5/2}}$ ($|3\rangle $), and $54{P_{3/2}}$ ($|4\rangle $). A probe laser ${\Omega _p}$ with a wavelength of ${\lambda _p} \sim 780$ nm and a coupling laser ${\Omega _c}$with ${\lambda _c} \sim 480$ nm counter-propagate through the atoms and drive $|1\rangle  \leftrightarrow |2\rangle $ and $|2\rangle  \leftrightarrow |3\rangle $ transition, respectively. A MW field drives the Rydberg transition of the states $|3\rangle $ to $|4\rangle $. A similar system has been used in  intracavity EIT \cite{peng2018cavity}, THz field measurement \cite{zhi2021terahertz}, nonlinear optical effects \cite{zhao2020enhancing}, and so on.

In the interaction picture and after the rotating wave approximation, the Hamiltonian of the system can be written as
\begin{equation}
\begin{array}{*{20}{l}}
H & =  - \hbar [{\Delta _p}|2\rangle \langle 2| + \left( {{\Delta _p} + {\Delta _c}} \right)|3\rangle \langle 3|\\ & \quad+ \left( {{\Delta _p} + {\Delta _c} - {\Delta _m}} \right)|4\rangle \langle 4| + {\Omega _p}|1\rangle \langle 2|
\\ & \quad+ {\Omega _c}|2\rangle \langle 3| + {\Omega _m}|3\rangle \langle 4| + {\rm{H}}.{\rm{C}}.],
\end{array}
\end{equation}
where \textit{H} is the interaction Hamiltonian, the Rabi frequency is ${\Omega _p} = {\mu _{12}}{E_p}/\hbar $, ${\Omega _c} = {\mu _{23}}{E_c}/\hbar $, and ${\Omega _m} = {\mu _{34}}{E_m}/\hbar $. ${\Delta _p} = {\omega _p} - {\omega _{12}}$, ${\Delta _c} = {\omega _c} - {\omega _{23}}$, and ${\Delta _m} = {\omega _m} - {\omega _{34}}$ denote the detunings of the corresponding fields, respectively. ${\mu _{ij}}$ (\textit{i}, \textit{j} = 1, 2, 3, 4) is the transition dipole moment from state $|i\rangle $ to state $|j\rangle $. The dynamic evolution of the system can be described using the density-matrix method as follows \cite{scully1997quantum}:
\begin{equation}
\dot \rho  =  - \frac{i}{\hbar }[H, \rho ] + L\left( \rho  \right),
\end{equation}
where ${L\left( \rho  \right)}$ denotes the decoherence processes. The time evolution of density matrix elements can be written as
\begin{flalign}
\begin{split}
&{\dot \rho _{11}} = {\Gamma _2}{\rho _{22}} - i\left( {{\rho _{12}} - {\rho _{21}}} \right){\Omega _p},\\
&{\dot \rho _{22}} =  - {\Gamma _2}{\rho _{22}} + {\Gamma _3}{\rho _{33}} - i[({\rho _{23}} - {\rho _{32}}){\Omega _c} + ({\rho _{12}} - {\rho _{21}}){\Omega _p}],\\
&{\dot \rho _{33}} =  - {\Gamma _3}{\rho _{33}} + {\Gamma _4}{\rho _{44}} - i[({\rho _{32}} - {\rho _{23}}){\Omega _c} + ({\rho _{34}} - {\rho _{43}}){\Omega _m}],\\
&{\dot \rho _{44}} =  - {\Gamma _4}{\rho _{44}} - i\left( {{\rho _{43}} - {\rho _{34}}} \right){\Omega _m},\\
&{\dot \rho _{21}} =  - \frac{1}{2}{\Gamma _2}{\rho _{21}} - i[{\Delta _p}{\rho _{21}} - {\rho _{31}}{\Omega _c} - ({\rho _{11}} + {\rho _{22}}){\Omega _p}],\\
 &{\dot \rho _{31}} =  - \frac{1}{2}{\Gamma _3}{\rho _{31}} - i[{\Delta _1}{\rho _{31}} - {\rho _{21}}{\Omega _c} - {\rho _{41}}{\Omega _m} + {\rho _{32}}{\Omega _p}],\\
&{\dot \rho _{41}} =  - \frac{1}{2}{\Gamma _4}{\rho _{41}} - i({\Delta _2}{\rho _{41}} - {\rho _{31}}{\Omega _m} + {\rho _{42}}{\Omega _p}),\\
 &{\dot \rho _{32}} =  - {\gamma _{32}}{\rho _{32}} - i[{\Delta _c}{\rho _{32}} + ({\rho _{33}} - {\rho _{22}}){\Omega _c} - {\rho _{42}}{\Omega _m} + {\rho _{31}}{\Omega _p}],\\
 &{\dot \rho _{42}} =  - {\gamma _{42}}{\rho _{42}} - i({\Delta _3}{\rho _{42}} + {\rho _{43}}{\Omega _c} - {\rho _{32}}{\Omega _m} + {\rho _{41}}{\Omega _p}),\\
 &{\dot \rho _{43}} =  - {\gamma _{43}}{\rho _{43}} - i[ - {\Delta _m}{\rho _{43}} + {\rho _{42}}{\Omega _c} + ({\rho _{44}} - {\rho _{33}}){\Omega _m}],
\end{split}&
\end{flalign}
with ${\rho _{ij}}{\rm{ = }}{\rho ^ * }_{ji}$ and the closure relation $\sum\nolimits_j {{\rho _{jj}}}  = 1$, (\textit{i},\textit{j}=1,2,3,4). Here, ${\Delta _1} = {\Delta _p} + {\Delta _c}$, ${\Delta _2} = {\Delta _p} + {\Delta _c} - {\Delta _m}$, and ${\Delta _3} = {\Delta _c} - {\Delta _m}$. ${\gamma _{ij}}$ is the decay from the states $|i\rangle $ to $|j\rangle $, and ${\gamma _{ij}} = ({\Gamma _i} + {\Gamma _j})/2$, ${\Gamma _i} $ is the population decay rate. We consider $\rho _{11}^{(0)} = 1$, $\rho _{ij}^{(0)} = 0$. The coherence term ${\rho _{21}}$ can be obtained by solving the above equation. With consideration of the residual Doppler effect, the frequency detunings of the control and probe fields are modified as ${{\rm{\delta }}_c} = {\Delta _c} - {k_c}v$, ${{\rm{\delta }}_p} = {\Delta _p} + {k_p}v$, where${k_c} = 2\pi /{\lambda _c}$, ${k_p} = 2\pi /{\lambda _p}$, and the susceptibility of the Rydberg atoms is then Doppler-averaged:
\begin{equation}
\chi {\rm{ = }}\frac{{2N{{\left| {{\mu _{12}}} \right|}^2}}}{{\hbar {\varepsilon _0}{\Omega _p}}}\frac{1}{{\sqrt \pi  u}}\int_{ - \infty }^{ + \infty } {{\rho _{21}}} {e^{ - \frac{{{v^2}}}{{{u^2}}}}}dv,\tag{4}
\end{equation}
where\textit{ N} is Rydberg atom density. ${{\mu _{12}}}$ is the dipole moment of transition $|1\rangle  \leftrightarrow |2\rangle $, ${{\varepsilon _0}}$ is the dielectric constant of vacuum, $u = \sqrt {2{k_B}T/m} $ is the most probable velocity of the atoms, ${{k_B}}$ is the Boltzmann constant, \textit{T} is the temperature of system, \textit{m} is the mass of the atom. Then, we obtain the three-photon coherence term in ${\rho _{21}}$ \cite{noh2011discrimination,noh2011diagrammatic}, and the three-photon coherence part of the atomic susceptibility is
\begin{equation}
\begin{array}{*{20}{l}}
{\chi _{{\rm{TPC}}}} & {\rm{ = }}\frac{{2N{{\left| {{\mu _{12}}} \right|}^2}}}{{\hbar {\varepsilon _0}{\Omega _p}}}\frac{1}{{\sqrt \pi  u}}\int_{ - \infty }^{ + \infty } {\frac{{i\Omega _m^2{\Omega _c}}}{{{C_5}{C_6}}}\frac{1}{{{C_7}}}} \\
 &  \times \frac{{i{\Omega _m}[{C_2}\Omega _c^2 + {C_1}( - {C_2}{C_3} - \Omega _m^2)] + {C_1}^2\Omega _m^2\Omega _p^3{\Omega _c}}}{{{C_1}{\Omega _m}\Omega _p^2({C_2} - {C_4}){C_8} + {C_9}}}dv,\tag{5}
\end{array}
\end{equation}
where ${C_1} = i{\Delta _m} - {G_{34}}$, ${C_2} = {G_{23}} + i{\delta _c}$, ${C_3} = {G_{24}} + i({\delta _c} - {\Delta _m})$, ${C_4} = {\Gamma _4}/2 + i({\delta _c} + {\delta _p} - {\Delta _m})$, ${C_5} = {\Gamma _2}/2 + i{\delta _p}$, ${C_6} = {\Gamma _3}/2 + i({\delta _c} + {\delta _p})$, ${C_7}{\rm{ = }}[(i{C_1}{C_4}{C_5}{C_6} - \Omega _c^2){\Omega _m} + i{C_1}{C_5}{\Omega _m}(\Omega _p^2 - \Omega _m^2)]$, ${C_8}{\rm{ = }}[{C_1}({C_5}{C_6} - \Omega _c^2){\Omega _m} + {C_1}{C_5}({\Omega _m} - {C_5}\Omega _c^2)]$, 
${C_9}{\rm{ = }}[i{\Omega _m}({C_2}\Omega _c^2 - i{C_1}{C_2}{C_3} + {C_1}\Omega _m^2) + {C_1}\Omega _p^2]$, ${G_{34}} = ({\Gamma _3} + {\Gamma _4})/2$, and ${G_{24}} = ({\Gamma _2} + {\Gamma _4})/2$.
\begin{figure}[!htbp]
\centering
\begin{minipage}[h]{0.34\textwidth}
\label{TPEIA spectrum (a.u.)1}
\includegraphics[width=5.9cm,height = 4.5cm]{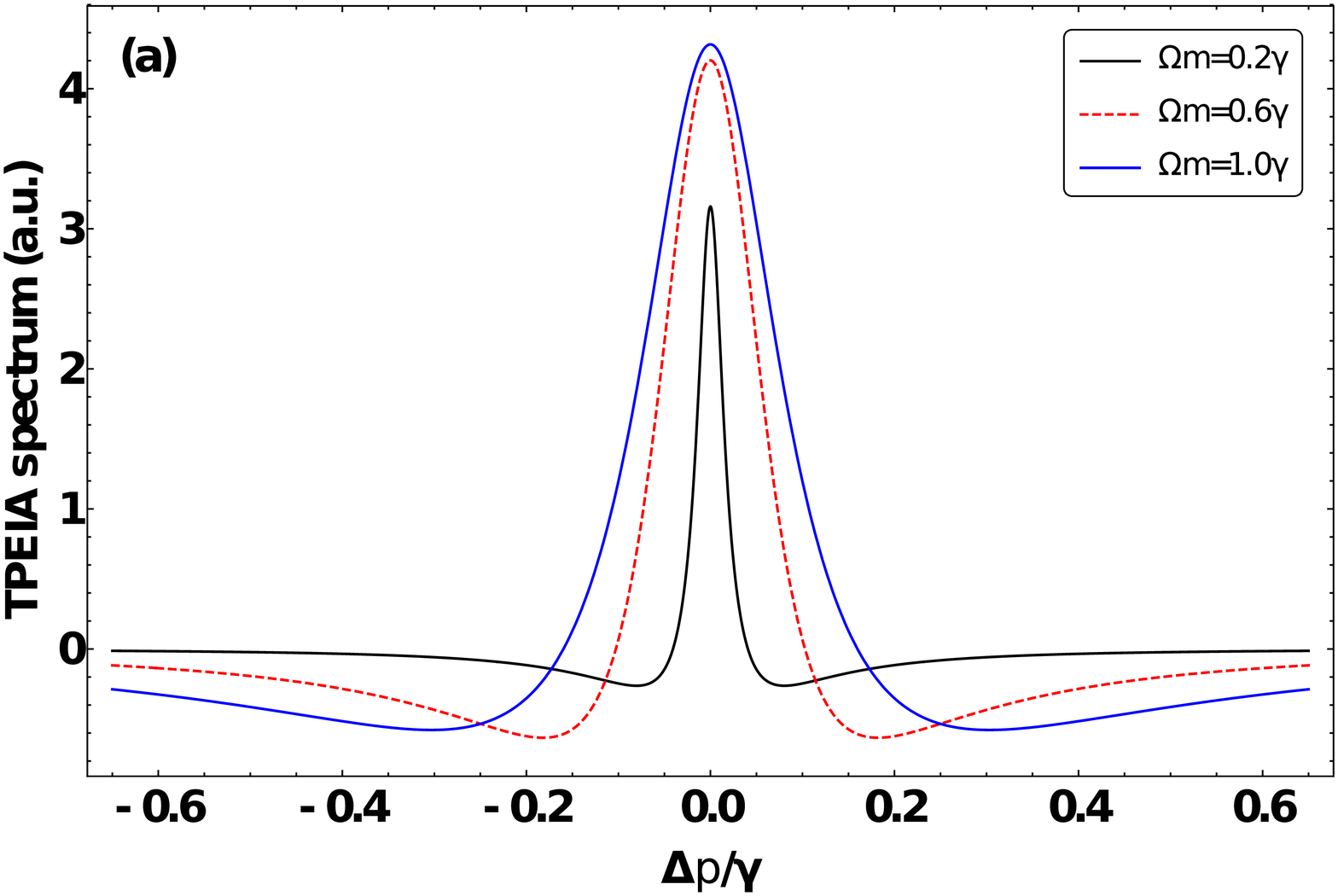}
\end{minipage}
\label{Fitting curve 1}
\includegraphics[width=6.8cm,height = 4.9cm]{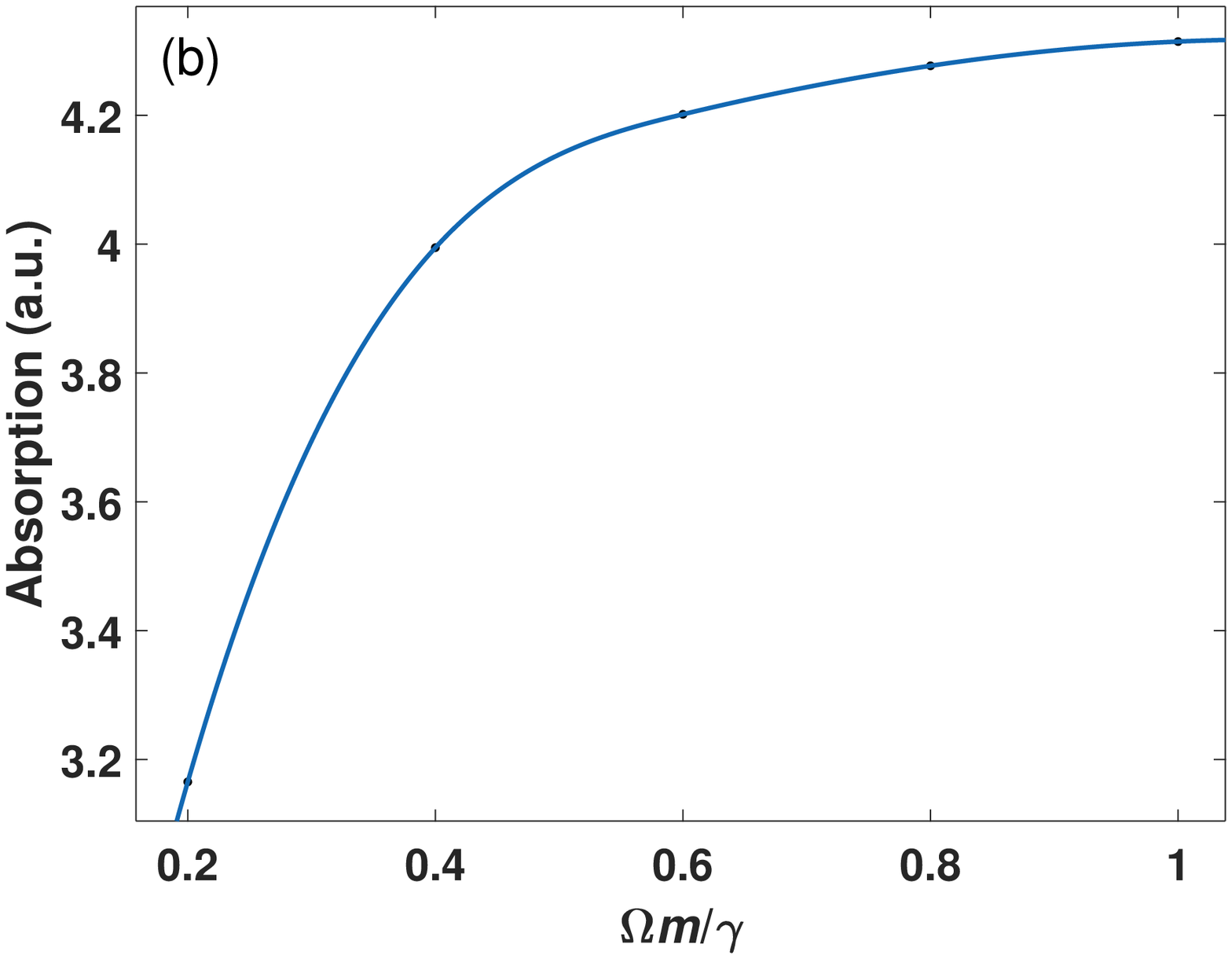}
\caption{(a) The TPEIA spectrum and (b) peak intensity as a function of the MW field strength, with ${\Omega _p} = 0.001\gamma $, ${\Omega _c} = 3{\rm{MHz}}$, ${\Gamma _3}{\rm{ = }}0.2\gamma $, ${\Gamma _4}{\rm{ = }}0.01\gamma $. ($\gamma  = 2\pi  \times 1{\rm{MHz}}$).}
\label{Fig 2}
\end{figure}
\begin{figure}[t]
\centering
\label{TPEIA spectrum (a.u.)2}
\includegraphics[width=5.7cm,height = 4.7cm]{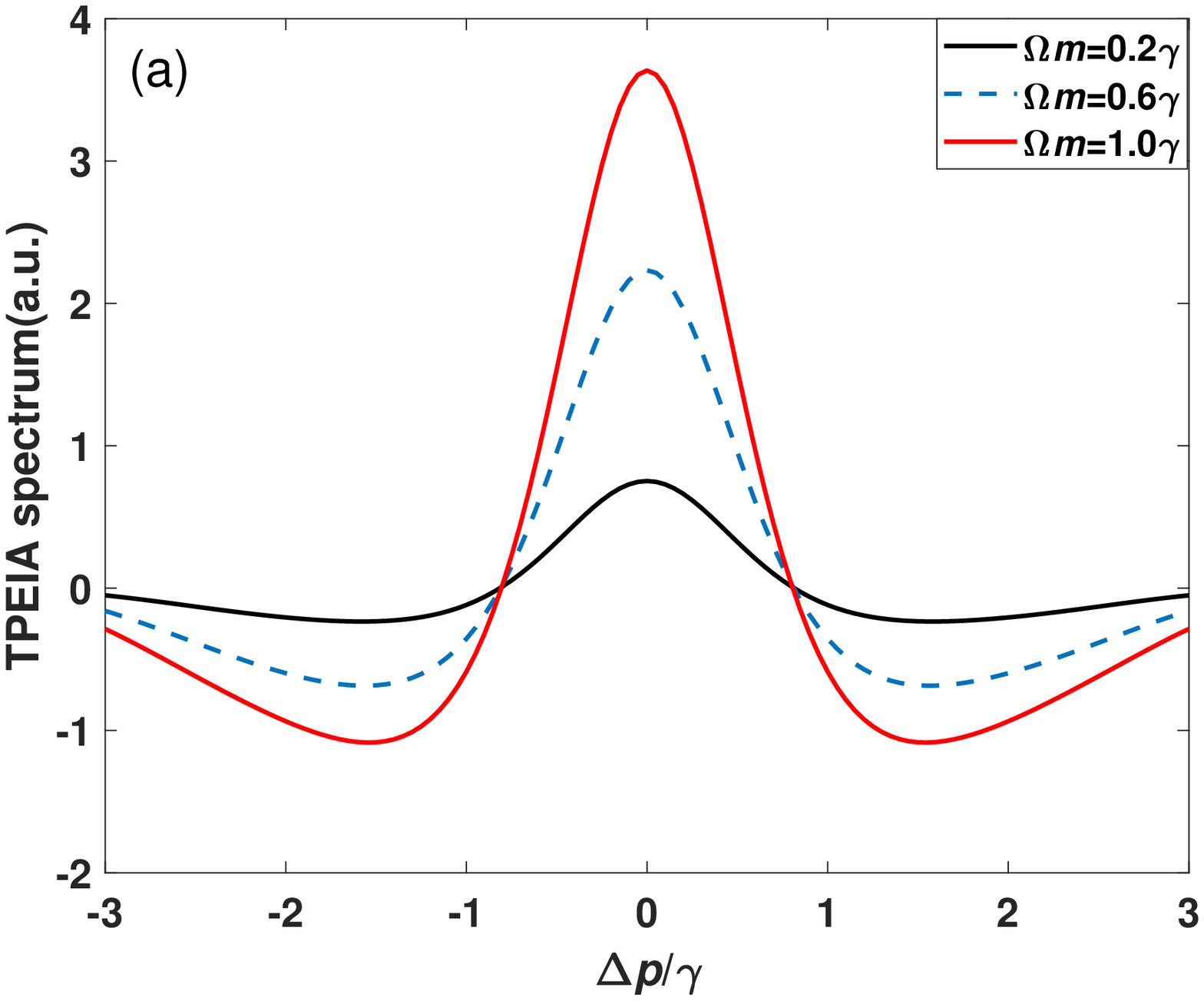}
\label{TPEIA spectrum (a.u.)3}
\includegraphics[width=5.7cm,height =4.7cm]{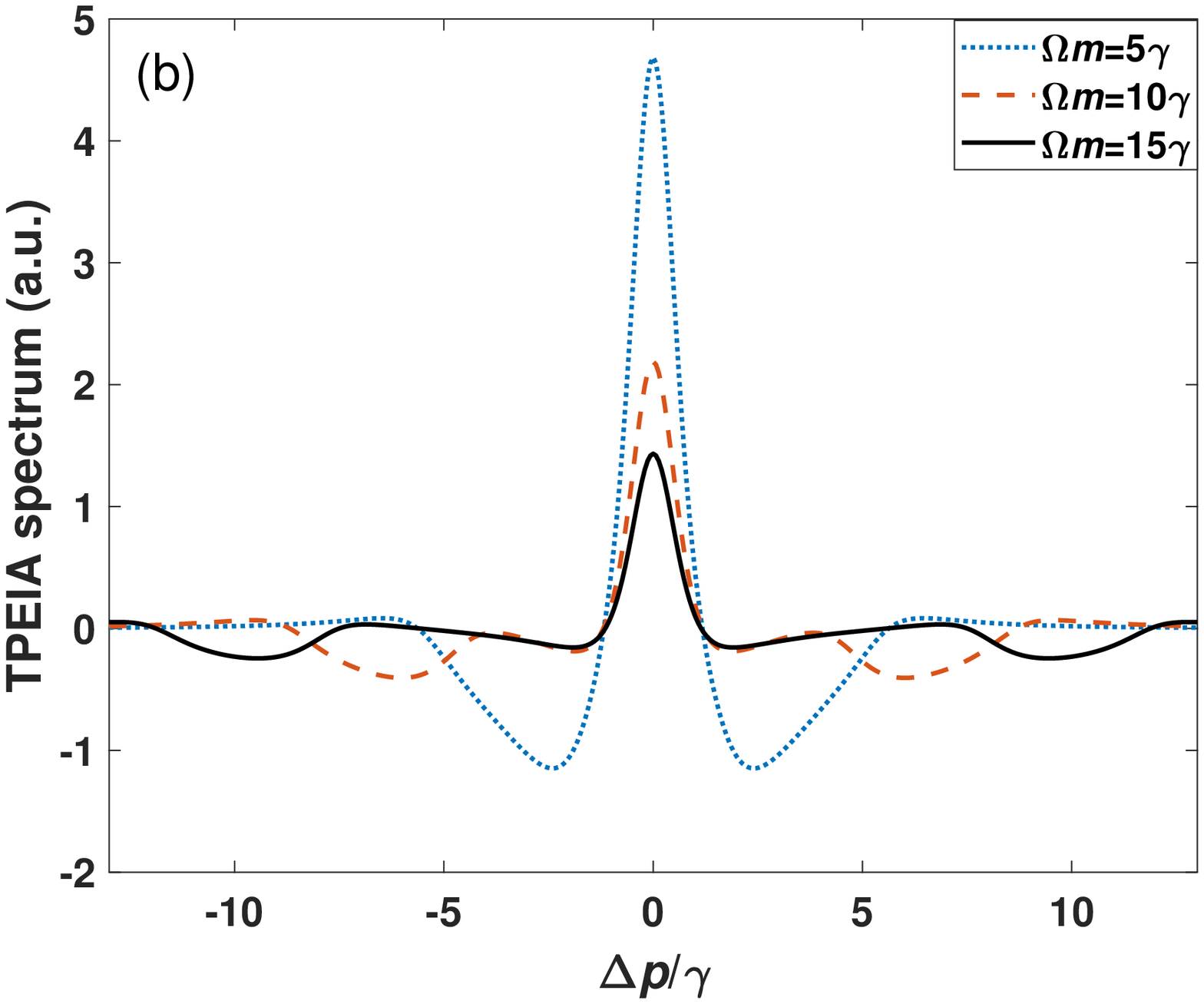}
\caption{(a) The Doppler-averaged TPEIA spectrum and (b) variation of the TPEIA spectrum with MW field. The other parameters are the same as in Fig. 2(a).}
\label{3}
\end{figure}
\section{RESULTS AND DISCUSSION}
We first consider the TPEIA spectrum changing with the MW field in a Doppler-free scheme, where the probe field counter-propagates with the control field (see Fig. 2). The ${}^{87}{\rm{Rb}}$ atomic density is $N = {10^{ - 7}}{\rm{c}}{{\rm{m}}^{ - 3}}$ and spontaneous decay ${\Gamma _2} = 6\gamma  = 2\pi  \times 6{\rm{MHz}}$ in the following discussion \cite{sedlacek2012microwave}. And the following discussion is scaled by $\gamma$ for simplicity. Fig. 2(a) shows that the TPEIA signal has one absorption peak through the three-photon process when the lasers interact with atoms resonantly. The basic property of TPEIA agree well with the previous study \cite{moon2014three,yadav2018study}, and here we pay attention to the variation of TPEIA with MW fields. For a weak MW field, the TPEIA peak increases with the strength of MW field, as shown in Fig. 2(a), the absorption peak becomes strong with increase of the MW field strength, and the peak linewidth becomes a little broad due to homogenous broadening effect. Thanks to the three-photon coherence, the population transfers from the ground state to the Rydberg state, and the peak value of absorption spectrum can be improved in the range of weak MW field. Fig. 2(b) shows the variations of magnitude of TPEIA peak as a function of the MW field strength. The TPEIA peak becomes strong by increasing MW field, and the linewidth remains narrow. While, the magnitude of TPEIA peak changes nonlinearly with the MW field strength, which may be not suitable for the linear measurement of MW electric field.
\begin{figure}[!htbp]
\centering
\label{Fitting curve 2}
\includegraphics[width=5.5cm,height = 4.7cm]{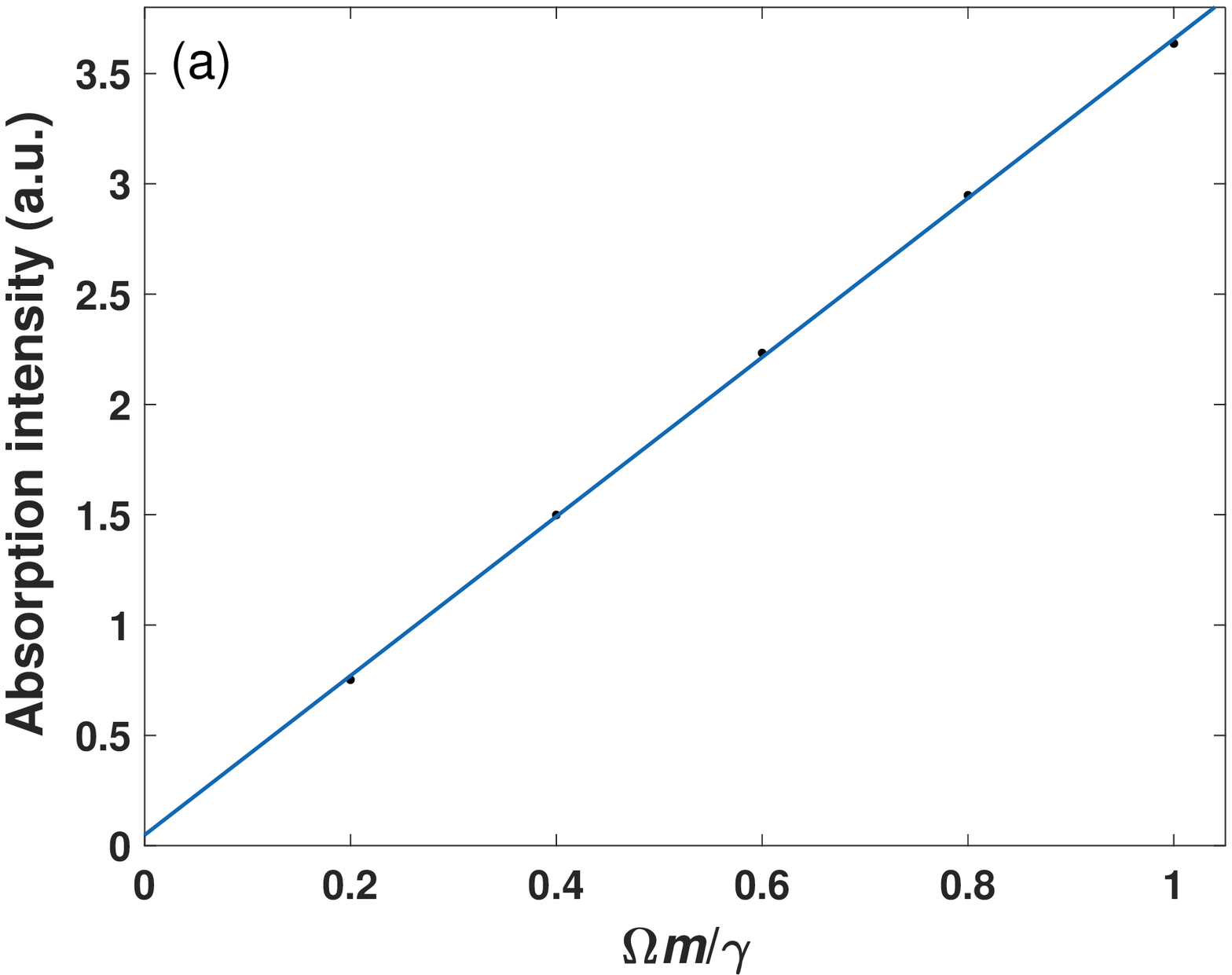}
\label{Fitting curve 3}
\includegraphics[width=5.5cm,height = 4.7cm]{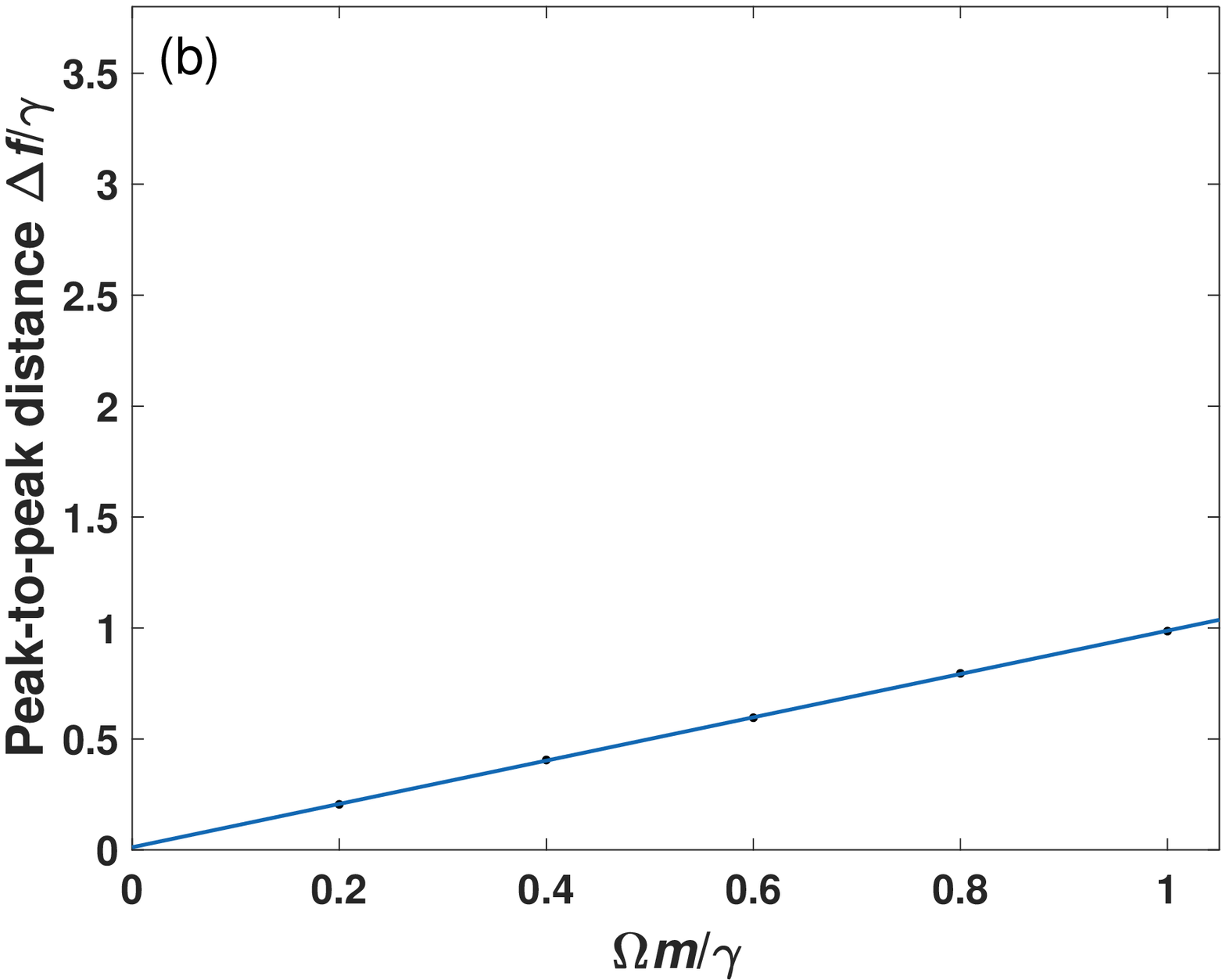}
\caption{(a) The variations of absorption peak intensity as a function of the MW field strength on the condition of residual Doppler effect and (b) Distance of two transmission peaks ${\Delta _f}$ versus the MW field strength. The other parameters are the same as in Fig. 2(a).}
\label{4}
\end{figure}
\begin{figure}[t]
\centering
\label{TPEIA spectrum (a.u.)4}
\includegraphics[width=5.7cm,height = 4.7cm]{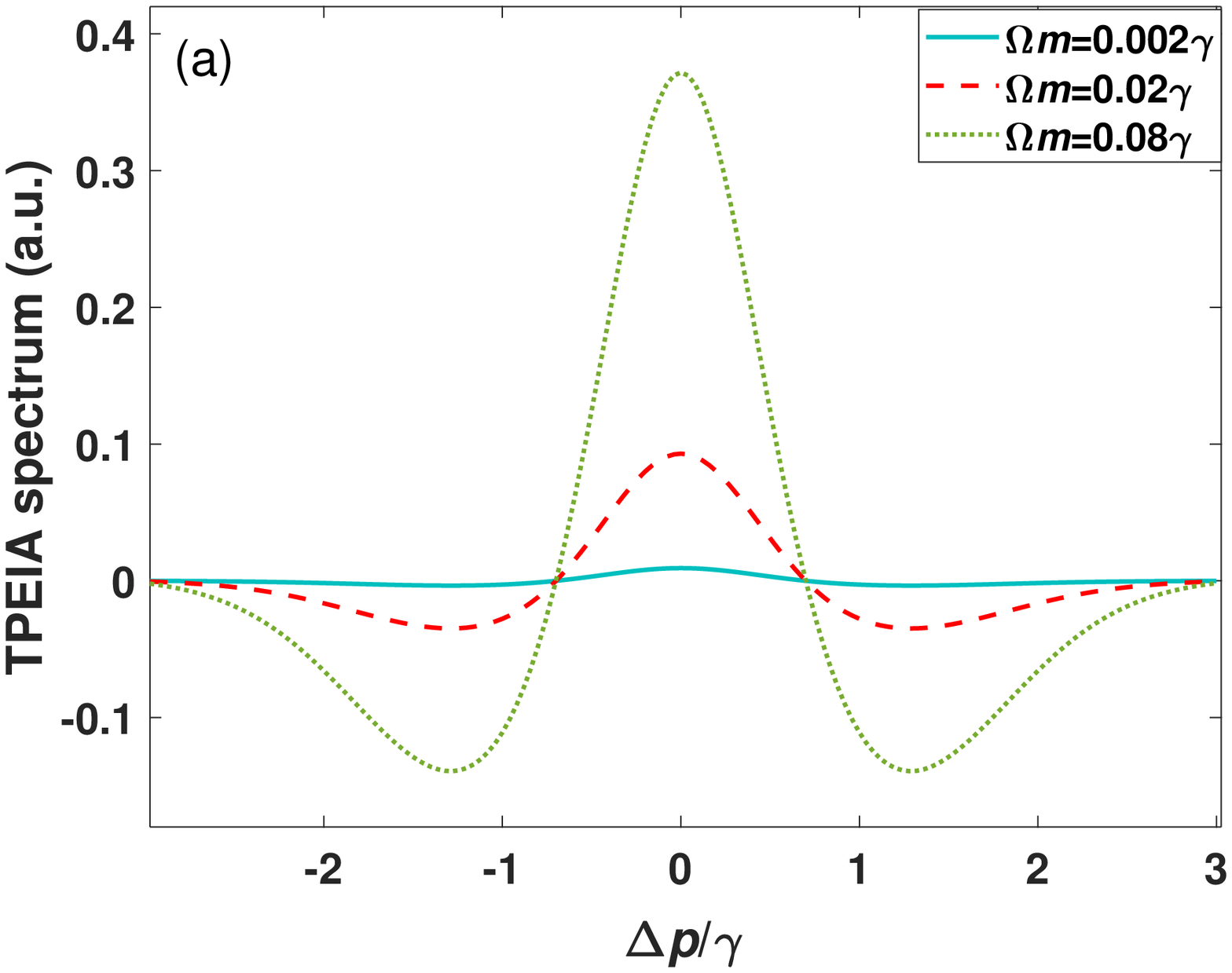}
\label{Transmission}
\includegraphics[width=5.7cm,height = 4.8cm]{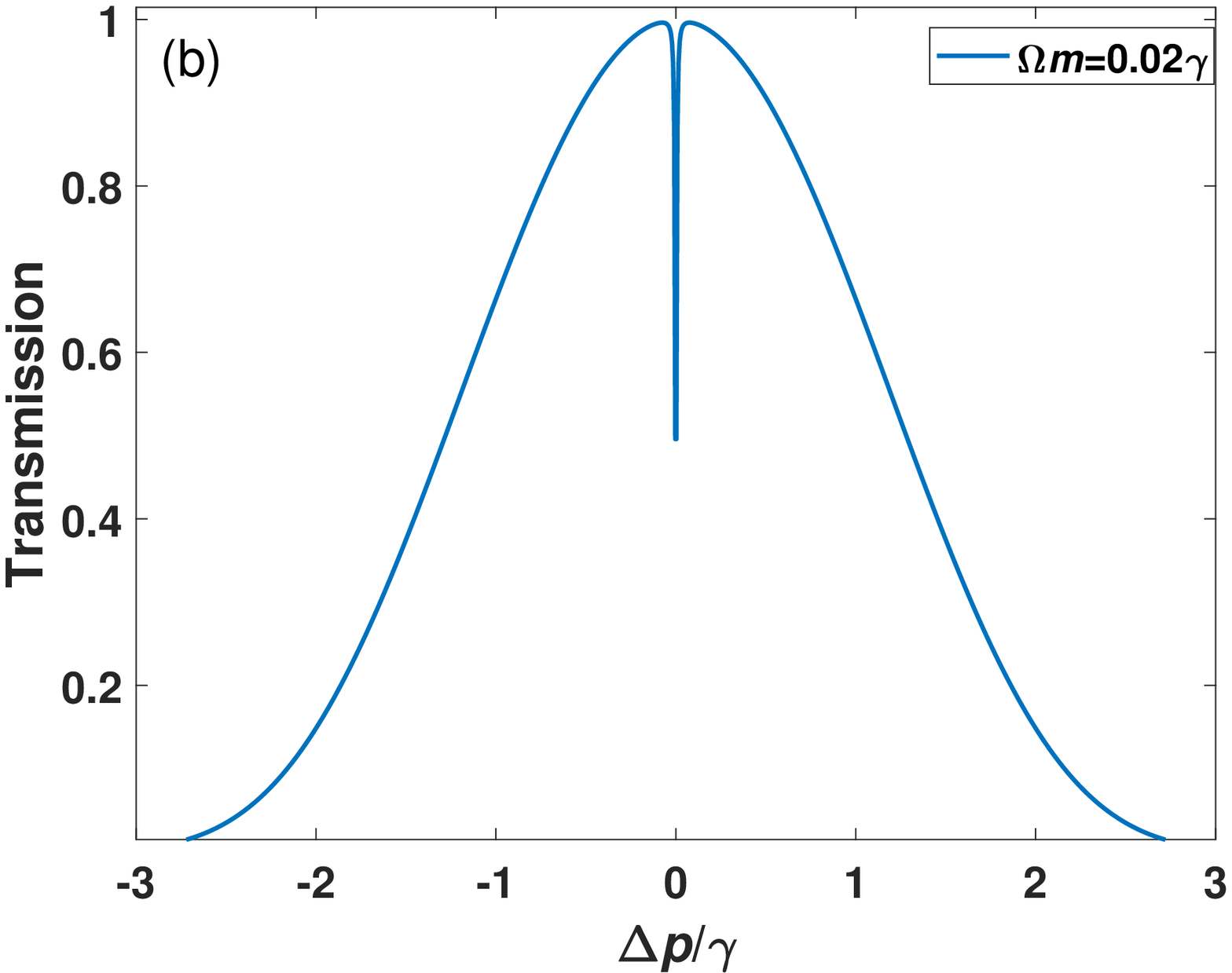}
\caption{(a) Doppler-averaged TPEIA spectrum and (b) transmission spectrum of EIT. The other parameters are the same as in Fig. 2(a).}
\label{5}
\end{figure}
Generally, most of the experiments are performed in a room temperature vapour. So, the Doppler effect is obvious due to mismatch of the coupling-field wavelength, where the probe and control fields counter-propagate through the atomic vapor. The variations of the absorption spectrum are shown in Fig. 3(a). The TPEIA peak becomes strong with increase of the MW field. However, when the strength of the MW field is further increases, the TPEIA signal is suppressed and two transmission windows far away from resonance (see Fig. 3(b)). We pay more attention to the enhanced TPEIA peak and explore its application in precise measurement. 

It is interesting to find that the magnitude of TPEIA peak varies linearly with the MW field, as shown in Fig. 4(a). The numerical results show that the curve slop based on three-photon coherence is about 4. Fig. 4(b) shows the linear measurement of MW field based on the common EIT method, where the frequency splitting of EIT peaks changes linearly with the MW electric field strength. The slope of measurement curve based on EIT is about 1 from simulation. The comparison of Fig. 4(a) and Fig. 4(b) shows that the curve slope based on TPEIA is about 4 times larger than that of EIT method. The  MW electric field strength could be estimated from changes in the magnitude of TPEIA peaks. It is known that, the larger curve slope under the same condition results in the better detection sensitivity. This indicates that the probe sensitivity could be improved by about a factor of 4 due to the three-photon coherence.

It is important to detect the minimum strength of the MW field for precise measurement. Fig. 5(a) shows the minimum detectable strength ${\rm{M}}{{\rm{W}}_{\min }}$ for the three-photon resonance case.  According to the Rayleigh criterion \cite{smith2013field}, the the corresponding spectrum resolution is about 0.02$\gamma $, which means the minimum detectable strength of the MW field is about 0.02$\gamma $ based on the EIT scheme, as shown in Fig. 5(b). Our simulation results show that the minimum detectable strength of the MW field is about 0.002$\gamma $ for the TPEIA spectrum, which is about 1/10 that based on an common EIT effect (see Fig. 5(a)). This indicates that the minimum detectable strength could be improved by 10 times due to the three-photon coherence.
\begin{figure}[t]
\centering
\label{TPEIA spectrum (a.u.)5}
\includegraphics[width=5.7cm,height = 4.7cm]{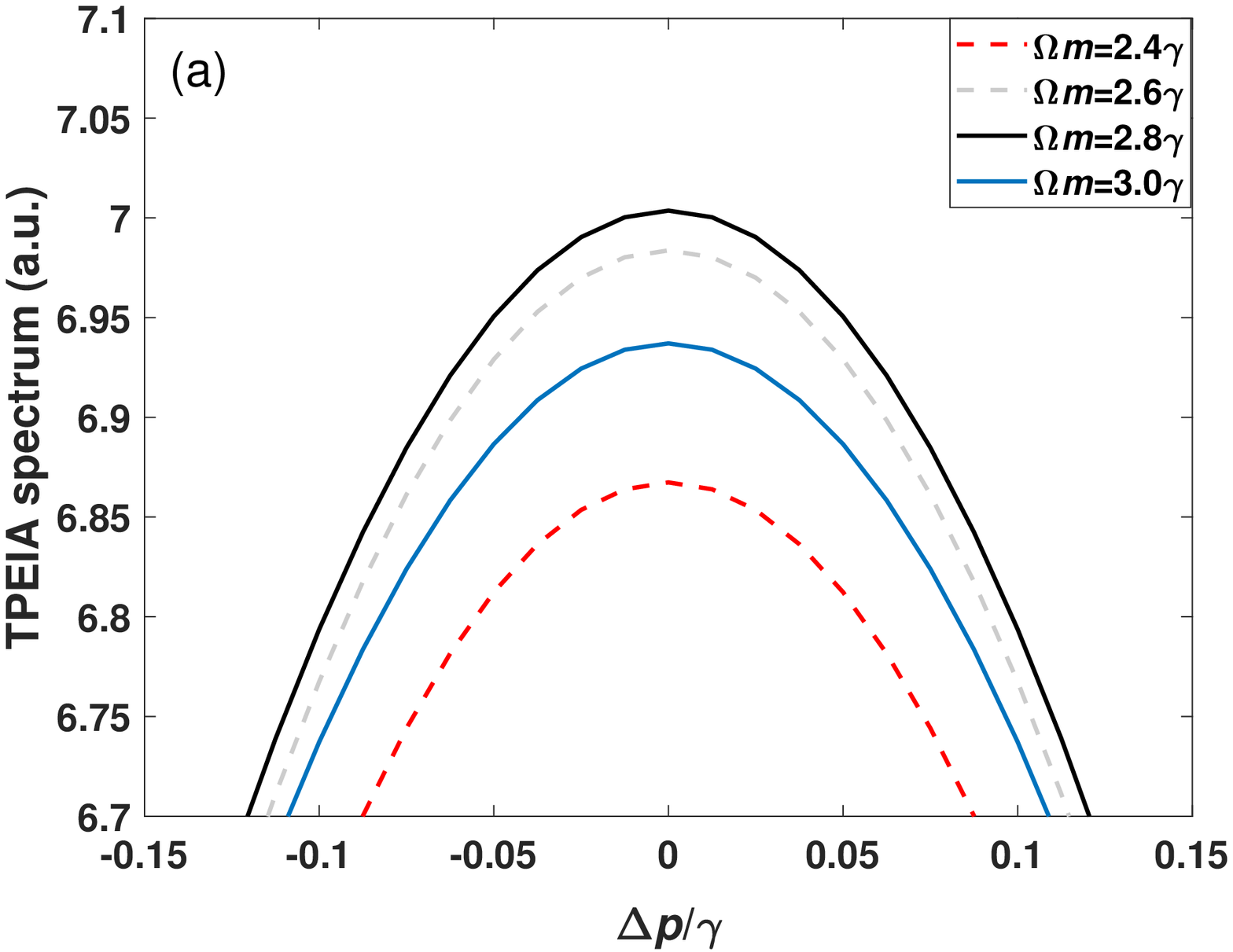}
\label{TPEIA spectrum (a.u.)6}
\includegraphics[width=5.8cm,height = 4.7cm]{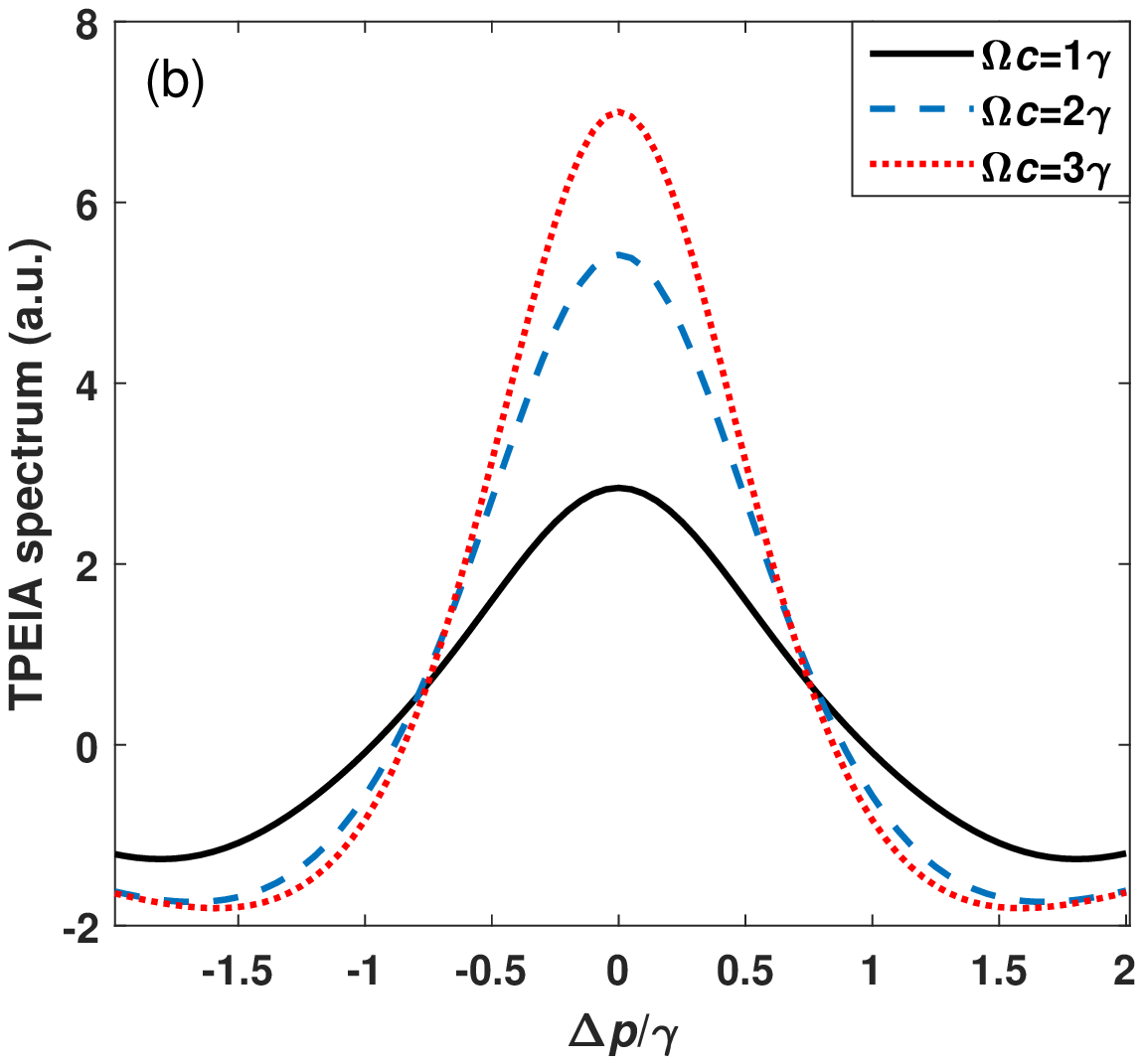}
\caption{(a) The effect of large MW field on the intensity of the Doppler-averaged TPEIA spectrum and (b) peak intensity as a function of the control field strength with ${\Omega _m} = 2.8\gamma $. The other parameters are the same as in Fig. 2(a).}
\label{6}
\end{figure}
\begin{figure}[!htbp]
\centering
\includegraphics[width=5.7cm,height = 4.7cm]{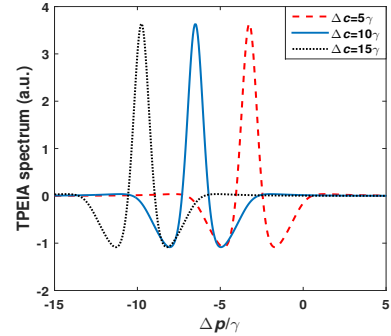}
\caption{Doppler-averaged TPEIA spectrum for different control field detuning ${\Delta _c}$, with ${\Omega _m} = \gamma $. The other parameters are the same as in Fig. 2(a).}
\label{7}
\end{figure}
The above discussions deal with the weak MW field. In the dressed-state picture, the probe and control transition consist of a classical Rydberg EIT scheme. There is an EIT window around the resonant frequency, and it can be understood in the EIT theory \cite{fleischhauer2005electromagnetically}. The coupling fields ${\Omega _c}$ dresses the states $|2\rangle $ and $|3\rangle $, and two new eigenstates appear, i.e., $| + \rangle $ and $| - \rangle $. With coupling of the probe field, two transition channels appear, $|1\rangle  \to | + \rangle $ and $|1\rangle  \to | - \rangle $. When the MW field drives the Rydberg transition $|3\rangle  \to |4\rangle $, the Rydberg-EIT is disturbed and an enhanced absorption peak builds up, which is referred to be as TPEIA.

Fig. 6(a) shows the effect of the large MW field on the magnitude of the Doppler-averaged TPEIA spectrum. For example, when the MW field strength ${\Omega _m} \le 2.8\gamma $, the TPEIA peak becomes strong with increase of the MW field, as shown Fig. 6(a). While, the TPEIA peak decreases with the further increase of the MW field. So, the TPEIA peak reaches a maximum at ${\Omega _m} = 2.8\gamma $ under the given condition. This is because the two dressed states induced by the MW field are well separated, and the AT splitting of Rydberg EIT is dominant over TPEIA peak in the regime of the large MW field \cite{kwak2016microwave}. The constructive interference for three-photon coherence gradually weakens. When the TPEIA peak decreases with ${\Omega _m}$, the peak-to-peak distance of the two transmission peaks increases with ${\Omega _m}$, as shown in Fig. 3(a). This effect can be well-understood in the dressed-state picture \cite{zhou2022three}. The effect of the control field on the intensity of the absorption spectrum is shown in Fig. 6(b). The magnitude of TPEIA peak increases when the control field becomes strong. The strong control field induces the good multi-photon coherence and contributes to the large TPEIA peak. Of course, if the control field is too large, the Rydberg EIT evolves into Aulter-Tonwns splitting, and the inter-path interference weakens, resulting in decrease of TPEIA peak.

In addition, the numerical results show that the linewidth of the absorption spectrum is about 1.15$\gamma $. The linewidth of TPEIA peak is a little broader than that of Doppler-free scheme. This is due to the residual Doppler effect as a result of the temperature of the atomic vapor. The above discussions are based on resonant interaction of the control field. Fig. 7 shows the effect of the control field detuning on the TPEIA spectrum. The figure shows that the absorption window shifts with the control field detuning ${\Delta _c}$. While the width and peak value of the TPEIA spectrum basically remain unchanged. This means that the scheme has a broad detection range and some tunability.

\section{Conclusion}
In summary, we theoretically investigate TPEIA spectrum of Rydberg atoms and propose to use three-photon coherence to detect the weak MW electric field. Due to the multi-photon coherence, there  is constructive interference in the TPEIA at the resonant frequency. It is interesting to find that the magnitude of TPEIA peaks change linearly with the MW field, which can be used to detect the MW electric field. The numerical results show the sensitivity based on TPEIA is about 4 times larger than that of the EIT scheme. Its minimum detectable strength is about one order of magnitude smaller than that of the EIT scheme. Moreover, the MW measurement based on TPEIA shows a broad detection range and some tunability. The proposed scheme may help to design novel MW-sensing devices.


%

\bibliographystyle{IEEEtran}

\bibliography{ref}

\begin{thebibliography}{10}
\providecommand{\url}[1]{#1}
\csname url@samestyle\endcsname
\providecommand{\newblock}{\relax}
\providecommand{\bibinfo}[2]{#2}
\providecommand{\BIBentrySTDinterwordspacing}{\spaceskip=0pt\relax}
\providecommand{\BIBentryALTinterwordstretchfactor}{4}
\providecommand{\BIBentryALTinterwordspacing}{\spaceskip=\fontdimen2\font plus
\BIBentryALTinterwordstretchfactor\fontdimen3\font minus
  \fontdimen4\font\relax}
\providecommand{\BIBforeignlanguage}[2]{{%
\expandafter\ifx\csname l@#1\endcsname\relax
\typeout{** WARNING: IEEEtran.bst: No hyphenation pattern has been}%
\typeout{** loaded for the language `#1'. Using the pattern for}%
\typeout{** the default language instead.}%
\else
\language=\csname l@#1\endcsname
\fi
#2}}
\providecommand{\BIBdecl}{\relax}
\BIBdecl

\bibitem{gallagher2005rydberg}
T.~Gallagher, \emph{Rydberg Atoms}.\hskip 1em plus 0.5em minus 0.4em\relax
  Cambridge, U.K.: Cambridge University, 2005.

\bibitem{boller1991observation}
K.~J. Boller, A.~Imamo{\u{g}}lu, and S.~E. Harris, ``Observation of
  electromagnetically induced transparency,'' \emph{Phys. Rev. Lett.}, vol.~66,
  no.~20, p. 2593, 1991.

\bibitem{alzetta1997induced}
G.~Alzetta, ``Induced transparency,'' \emph{Phys. Today}, vol.~50, no.~7, pp.
  36--42, 1997.

\bibitem{sedlacek2012microwave}
J.~A. Sedlacek, A.~Schwettmann, H.~K{\"u}bler, R.~L{\"o}w, T.~Pfau, and J.~P.
  Shaffer, ``Microwave electrometry with {Rydberg} atoms in a vapour cell using
  bright atomic resonances,'' \emph{Nat. Phys.}, vol.~8, no.~11, pp. 819--824,
  2012.

\bibitem{cheng2017high}
H.~Cheng, H.~Wang, S.~Zhang, P.~Xin, J.~Luo, and H.~Liu, ``High resolution
  electromagnetically induced transparency spectroscopy of {Rydberg} 87 {Rb}
  atom in a magnetic field,'' \emph{Opt. Exp.}, vol.~25, no.~26, pp.
  33\,575--33\,587, 2017.

\bibitem{simons2016using}
M.~T. Simons, J.~A. Gordon, C.~L. Holloway, D.~A. Anderson, S.~A. Miller, and
  G.~Raithel, ``Using frequency detuning to improve the sensitivity of electric
  field measurements via electromagnetically induced transparency and
  {Autler}-{Townes} splitting in {Rydberg} atoms,'' \emph{Appl. Phys. Lett.},
  vol. 108, no.~17, p. 174101, 2016.

\bibitem{fan2015atom}
H.~Fan, S.~Kumar, J.~Sedlacek, H.~K{\"u}bler, S.~Karimkashi, and J.~P. Shaffer,
  ``Atom based {RF} electric field sensing,'' \emph{J. Phys. B: At. Mol. Opt.
  Phys.}, vol.~48, no.~20, p. 202001, 2015.

\bibitem{sedlacek2013atom}
J.~Sedlacek, A.~Schwettmann, H.~K{\"u}bler, and J.~Shaffer, ``Atom-based vector
  microwave electrometry using rubidium {Rydberg} atoms in a vapor cell,''
  \emph{Phys. Rev. Lett.}, vol. 111, no.~6, p. 063001, 2013.

\bibitem{jing2020atomic}
M.~Jing, Y.~Hu, J.~Ma, H.~Zhang, L.~Zhang, L.~Xiao, and S.~Jia, ``Atomic
  superheterodyne receiver based on microwave-dressed {Rydberg} spectroscopy,''
  \emph{Nat. Phys.}, vol.~16, no.~9, pp. 911--915, 2020.

\bibitem{prajapati2021enhancement}
N.~Prajapati, A.~K. Robinson, S.~Berweger, M.~T. Simons, A.~B. Artusio-Glimpse,
  and C.~L. Holloway, ``Enhancement of electromagnetically induced transparency
  based {Rydberg}-atom electrometry through population repumping,'' \emph{Appl.
  Phys. Lett.}, vol. 119, no.~21, p. 214001, 2021.

\bibitem{simons2021continuous}
M.~T. Simons, A.~B. Artusio-Glimpse, C.~L. Holloway, E.~Imhof, S.~R. Jefferts,
  R.~Wyllie, B.~C. Sawyer, and T.~G. Walker, ``Continuous radio-frequency
  electric-field detection through adjacent {Rydberg} resonance tuning,''
  \emph{Phys. Rev. A}, vol. 104, no.~3, p. 032824, 2021.

\bibitem{jia2021span}
F.~D. Jia, X.~B. Liu, J.~Mei, Y.~H. Yu, H.~Y. Zhang, Z.~Q. Lin, H.~Y. Dong,
  J.~Zhang, F.~Xie, and Z.~P. Zhong, ``Span shift and extension of quantum
  microwave electrometry with {Rydberg} atoms dressed by an auxiliary microwave
  field,'' \emph{Phys. Rev. A}, vol. 103, no.~6, p. 063113, 2021.

\bibitem{simons2019rydberg}
M.~T. Simons, A.~H. Haddab, J.~A. Gordon, and C.~L. Holloway, ``A {Rydberg}
  atom-based mixer: Measuring the phase of a radio frequency wave,''
  \emph{Appl. Phys. Lett.}, vol. 114, no.~11, p. 114101, 2019.

\bibitem{lin2022sensitive}
L.~Lin, Y.~He, Z.~Yin, D.~Li, Z.~Jia, Y.~Zhao, B.~Chen, and Y.~Peng,
  ``Sensitive detection of radio-frequency field phase with interacting dark
  states in {Rydberg} atoms,'' \emph{Appl. Opt.}, vol.~61, no.~6, pp.
  1427--1433, 2022.

\bibitem{mcgloin2001electromagnetically}
D.~McGloin, D.~Fulton, and M.~Dunn, ``Electromagnetically induced transparency
  in {N}-level cascade schemes,'' \emph{Opt. Commun.}, vol. 190, no. 1-6, pp.
  221--229, 2001.

\bibitem{carr2012three}
C.~Carr, M.~Tanasittikosol, A.~Sargsyan, D.~Sarkisyan, C.~S. Adams, and K.~J.
  Weatherill, ``Three-photon electromagnetically induced transparency using
  {Rydberg} states,'' \emph{Opt. Lett.}, vol.~37, no.~18, pp. 3858--3860, 2012.

\bibitem{moon2014three}
H.~S. Moon and T.~Jeong, ``Three-photon electromagnetically induced absorption
  in a ladder-type atomic system,'' \emph{Phys. Rev. A}, vol.~89, no.~3, p.
  033822, 2014.

\bibitem{lee2015relationship}
Y.~S. Lee, H.~R. Noh, and H.~S. Moon, ``Relationship between two-and
  three-photon coherence in a ladder-type atomic system,'' \emph{Opt. Exp.},
  vol.~23, no.~3, pp. 2999--3009, 2015.

\bibitem{kwak2016microwave}
H.~M. Kwak, T.~Jeong, Y.~S. Lee, and H.~S. Moon, ``Microwave-induced
  three-photon coherence of {Rydberg} atomic states,'' \emph{Opt. Commun.},
  vol. 380, pp. 168--173, 2016.

\bibitem{shaffer2018read}
J.~Shaffer and H.~K{\"u}bler, ``A read-out enhancement for microwave electric
  field sensing with {Rydberg} atoms,'' \emph{Proc. SPIE}, vol. 10674, p.
  106740C, 2018.

\bibitem{you2022microwave}
S.~You, M.~Cai, S.~Zhang, Z.~Xu, and H.~Liu, ``Microwave-field sensing via
  electromagnetically induced absorption of {Rb} irradiated by three-color
  infrared lasers,'' \emph{Opt. Exp.}, vol.~30, no.~10, pp. 16\,619--16\,629,
  2022.

\bibitem{bai2022autler}
J.~Bai, Y.~Jiao, Y.~He, R.~Song, J.~Zhao, and S.~Jia, ``Autler-townes splitting
  of three-photon excitation of cesium cold {Rydberg} gases,'' \emph{Opt.
  Exp.}, vol.~30, no.~10, pp. 16\,748--16\,757, 2022.

\bibitem{gutekunst2017fiber}
J.~Gutekunst, D.~Weller, H.~K{\"u}bler, J.-P. Negel, M.~A. Ahmed, T.~Graf, and
  R.~L{\"o}w, ``Fiber-integrated spectroscopy device for hot alkali vapor,''
  \emph{Appl. Opt.}, vol.~56, no.~21, pp. 5898--5902, 2017.

\bibitem{zhang2018thermal}
S.~Zhang, Y.~Hu, G.~Lin, Y.~Niu, K.~Xia, J.~Gong, and S.~Gong,
  ``Thermal-motion-induced non-reciprocal quantum optical system,'' \emph{Nat.
  Photon.}, vol.~12, no.~12, pp. 744--748, 2018.

\bibitem{ying2014realization}
K.~Ying, Y.~Niu, D.~Chen, H.~Cai, R.~Qu, and S.~Gong, ``Realization of cavity
  linewidth narrowing via interacting dark resonances in a tripod-type
  electromagnetically induced transparency system,'' \emph{J. Opt. Soc. Am. B},
  vol.~31, no.~1, pp. 144--148, 2014.

\bibitem{peng2018cavity}
Y.~Peng, J.~Wang, A.~Yang, Z.~Jia, D.~Li, and B.~Chen, ``Cavity-enhanced
  microwave electric field measurement using {Rydberg} atoms,'' \emph{J. Opt.
  Soc. Am. B}, vol.~35, no.~9, pp. 2272--2277, 2018.

\bibitem{zhi2021terahertz}
C.~Zhi~Wen, S.~Zhen~Yue, L.~Kai~Yu, H.~Wei, Y.~Hui, and Z.~Shi~Liang,
  ``Terahertz measurement based on {Rydberg} atomic antenna,'' \emph{Acta Phys.
  Sin.}, vol.~70, no.~6, 2021.

\bibitem{zhao2020enhancing}
S.~Zhao, W.~Zhou, Y.~Cai, Z.~Chang, Q.~Zeng, and Y.~Peng, ``Enhancing optical
  delay using cross-{Kerr} nonlinearity in {Rydberg} atoms,'' \emph{Appl.
  Opt.}, vol.~59, no.~32, pp. 10\,076--10\,081, 2020.

\bibitem{vogt2018microwave}
T.~Vogt, C.~Gross, T.~F. Gallagher, and W.~Li, ``Microwave-assisted {Rydberg}
  electromagnetically induced transparency,'' \emph{Opt. lett.}, vol.~43,
  no.~8, pp. 1822--1825, 2018.

\bibitem{scully1997quantum}
M.~O. Scully and M.~Zubairy, \emph{Quantum Optics}.\hskip 1em plus 0.5em minus
  0.4em\relax Cambridge, U.K.: Cambridge University, 1997.

\bibitem{noh2011discrimination}
H.~R. Noh and H.~S. Moon, ``Discrimination of one-photon and two-photon
  coherence parts in electromagnetically induced transparency for a ladder-type
  three-level atomic system,'' \emph{Opt. Exp.}, vol.~19, no.~12, pp.
  11\,128--11\,137, 2011.

\bibitem{noh2011diagrammatic}
H.~R. {}Noh and H.~S. Moon, ``Diagrammatic analysis of multiphoton processes in
  a ladder-type three-level atomic system.''

\bibitem{yadav2018study}
K.~Yadav and A.~Wasan, ``Study of coherence effects in a four-level {$\Xi$}-
  {$\Lambda$} type system,'' \emph{J. Phys. B: At. Mol. Opt. Phys.}, vol.~51,
  no.~10, p. 105501, 2018.

\bibitem{zhou2022three}
K.~Zhou, X.~a. Yan, Y.~Han, W.~Xu, H.~Ti, H.~Liu, and Y.~Chen, ``Three-photon
  electromagnetically induced absorption in a dressed atomic system,'' \emph{J.
  Opt. Soc. Am. B}, vol.~39, no.~2, pp. 501--507, 2022.

\bibitem{smith2013field}
D.~G. Smith, \emph{Field guide to physical optics}.\hskip 1em plus 0.5em minus
  0.4em\relax USA: SPIE Press, 2013, vol. FG17.

\bibitem{holloway2014broadband}
C.~L. Holloway, J.~A. Gordon, S.~Jefferts, A.~Schwarzkopf, D.~A. Anderson,
  S.~A. Miller, N.~Thaicharoen, and G.~Raithel, ``Broadband {Rydberg}
  atom-based electric-field probe for {SI}-traceable, self-calibrated
  measurements,'' \emph{IEEE Trans. Antennas Propagat}, vol.~62, no.~12, pp.
  6169--6182, 2014.

\bibitem{osterwalder1999using}
A.~Osterwalder and F.~Merkt, ``Using high {Rydberg} states as electric field
  sensors,'' \emph{Phys. Rev. Lett.}, vol.~82, no.~9, p. 1831, 1999.

\bibitem{ludlow2015optical}
A.~D. Ludlow, M.~M. Boyd, J.~Ye, E.~Peik, and P.~O. Schmidt, ``Optical atomic
  clocks,'' \emph{Rev. Mod. Phys}, vol.~87, no.~2, p. 637, 2015.

\bibitem{sun2017measuring}
F.~Sun, J.~Ma, Q.~Bai, X.~Huang, B.~Gao, and D.~Hou, ``Measuring microwave
  cavity response using atomic {Rabi} resonances,'' \emph{Appl. Phys. Lett.},
  vol. 111, no.~5, p. 051103, 2017.

\bibitem{kominis2003subfemtotesla}
I.~Kominis, T.~Kornack, J.~Allred, and M.~V. Romalis, ``A subfemtotesla
  multichannel atomic magnetometer,'' \emph{Nature}, vol. 422, no. 6932, pp.
  596--599, 2003.

\bibitem{kucsko2013nanometre}
G.~Kucsko, P.~C. Maurer, N.~Y. Yao, M.~Kubo, H.~J. Noh, P.~K. Lo, H.~Park, and
  M.~D. Lukin, ``Nanometre-scale thermometry in a living cell,'' \emph{Nature},
  vol. 500, no. 7460, pp. 54--58, 2013.

\bibitem{fleischhauer2005electromagnetically}
M.~Fleischhauer, A.~Imamoglu, and J.~P. Marangos, ``Electromagnetically induced
  transparency: Optics in coherent media,'' \emph{Rev. Mod. Phys.}, vol.~77,
  no.~2, p. 633, 2005.

\bibitem{robinson2021determining}
A.~K. Robinson, N.~Prajapati, D.~Senic, M.~T. Simons, and C.~L. Holloway,
  ``Determining the angle-of-arrival of a radio-frequency source with a
  {Rydberg} atom-based sensor,'' \emph{Appl. Phys. Lett.}, vol. 118, no.~11, p.
  114001, 2021.

\bibitem{zhou2022theoretical}
Y.~Zhou, R.~Peng, J.~Zhang, L.~Zhang, Z.~Song, Z.~Feng, and Y.~Peng,
  ``Theoretical investigation on the mechanism and law of broadband terahertz
  wave detection using {Rydberg} quantum state,'' \emph{IEEE Photon. J.},
  vol.~14, no.~3, pp. 1--8, 2022.

\bibitem{tu2017angle}
Z.~Tu, A.~Wen, Z.~Xiu, W.~Zhang, and M.~Chen, ``Angle-of-arrival estimation of
  broadband microwave signals based on microwave photonic filtering,''
  \emph{IEEE Photon. J.}, vol.~9, no.~5, pp. 1--8, 2017.

\end{thebibliography}

\vfill

\end{document}